\documentclass{iaus}
\usepackage{graphicx}

  \checkfont{eurm10}
  \iffontfound
    \IfFileExists{upmath.sty}
      {\typeout{^^JFound AMS Euler Roman fonts on the system,
                   using the 'upmath' package.^^J}%
       \usepackage{upmath}}
      {\typeout{^^JFound AMS Euler Roman fonts on the system, but you
                   dont seem to have the}%
       \typeout{'upmath' package installed. iaus.cls can take advantage
                 of these fonts,^^Jif you use 'upmath' package.^^J}%
      }
  \else
  \fi

  \checkfont{msam10}
  \iffontfound
    \IfFileExists{amssymb.sty}
      {\typeout{^^JFound AMS Symbol fonts on the system, using the
                'amssymb' package.^^J}%
       \usepackage{amssymb}%

      }{}
  \fi

  \IfFileExists{amsbsy.sty}
    {\typeout{^^JFound the 'amsbsy' package on the system, using it.^^J}%
     \usepackage{amsbsy}}
    {\providecommand\boldsymbol[1]{\mbox{\boldmath $##1$}}}

\newsavebox{\astrutbox}
\sbox{\astrutbox}{\rule[-5pt]{0pt}{20pt}}

\newcommand\hmpc{~h$^{-1}$ Mpc~}

\title[Outskirts of Galaxy Clusters: intense life in the suburbs]
      {Interacting clusters and their environment}
\author[S. Bardelli]%
{S.Bardelli}
\affiliation{INAF-Osservatorio Astronomico di Bologna, via Ranzani 1,
40127 Bologna (Italy) \\
e-mail: sandro.bardelli@bo.astro.it
}

\pubyear{2004}
\volume{195}
\pagerange{1--8}
\date{?? and in revised form ??}
\jname{Outskirts of Galaxy Clusters: intense life in the suburbs}
\editors{A. Diaferio, ed.}
\begin{document}
\ifx\href\undefined\else\hypersetup{linktocpage=true}\fi
\maketitle

\begin{abstract}
Central regions of superclusters are the ideal places where to study
cluster merging phenomena: in fact the accretion activity is enhanced,
as predicted by the cosmological simulations. In this paper I review
the case-study of the Shapley Concentration, aimed to understand the effect of
major mergings on the intracluster medium and the galaxy population of the
involved clusters. 
\end{abstract}

\section{Introduction}

Cluster mergings are known to be among the most energetic phenomena
in the Universe, but until now studies at all wavelengths have not been 
extensively carried on:
therefore it is still unclear how the collision energy is dissipated and 
which is the effect of merging on the emission properties of the galaxies and 
on the physics of the intracluster medium.

In cosmological N-body simulations the cluster accretion happens along
specific directions defined by the density caustics and richer clusters
form preferentially where the environment density is higher. 
Superclusters can be considered as the observational counterparts of these
caustics and it is expected that at their centers the cluster accretion
is still strongly active. Therefore, superclusters are the ideal places 
where to study merging phenomena, because the cross-section for cluster 
collisions is enhanced.

The best place for these studies is the central region of the Shapley
Concentration supercluster, a huge concentration of clusters at $z\sim 0.05$
(\cite{ray89}, \cite{plionis91}, 
\cite[Raychaudhury et al. 1991]{raychaudhury91},
\cite[Zucca et al. 1993]{zucca93}). 
This region is anomalously rich in clusters, considering that it has 25 members 
while the Great Attractor (with a similar mass overdensity) contains only 
6 clusters (see Table 3 of \cite[Zucca et al. 1993]{zucca93}): for some 
reasons, in this region the cluster formation efficiency has been enhanced 
also with respect to similar regions. 
Moreover, as noted by \cite{raychaudhury91}, the fraction of clusters with 
substructures in this supercluster is higher than elsewhere, meaning that the 
process of cluster formation is still strongly active, suggesting that this 
region could be considered a ``nursery" of rich clusters.

\begin{figure}
\centering
\caption{Central region of the Shapley Concentration supercluster (figure 
from \cite[Drinkwater et al. 2004]{drinkwater04}; reproduced by permission of 
CSIRO Publishing, Melbourne, Australia; copyright Astronomical Society of 
Australia).
Note the two structures (cluster complexes) at $\alpha=13^h 30^m$ and 
$12^h 55^m$, connected by a bridge of galaxies 
resembling the Great Wall. Circles indicate the position of clusters.
}
\end{figure}

\section{Large scale structure}

Various redshift surveys have been conducted in order to study the
large scale distribution of galaxies in the Shapley Concentration 
(\cite{quintana00}, \cite[Bardelli et al. 2000]{bardelli00},
\cite[Drinkwater et al. 2004]{drinkwater04}) and its relation with the 
distribution of clusters.
Up to now, there are a few thousands of galaxy redshifts in the supercluster,
allowing not only to determine its geometry but also its overdensity and mass 
(see Figure 1). The distribution of inter-cluster galaxies is well
described by a plane tilted with respect to the line of sight: the distribution
of galaxies around this plane is a Gaussian with a dispersion of $3.8$ \hmpc
(\cite[Bardelli et al. 2000]{bardelli00}).
The huge overdensity in number of galaxies ($\sim 11$ on a scale of $\sim 10$ 
\hmpc) found for this supercluster is consistent only with $\Lambda$CDM or 
open CDM cosmological scenarios (\cite[Bardelli et al. 2000]{bardelli00};
for a more recent determination on larger scales see 
\cite[Drinkwater et al. 2004]{drinkwater04}).
The determination of the mass of the supercluster is more difficult because, 
as indicated by the overdensity, the region is far from virialization.
Various methods have been applied (\cite[Ettori et al. 1997]{ettori97}, 
\cite[Bardelli et al. 2000]{bardelli00}, 
\cite[Reisenegger et al. 2000]{reisen00}), leading to an estimate of 
$\sim 10^{16}\ M_{\odot}$.

All these properties are extreme also for $\Lambda$CDM models and for this 
reason it is quite difficult to find a supercluster like the Shapley
Concentration in numerical simulations.

As can be seen in Figure 2, two main groups of clusters (``cluster complexes") 
dominate the central region of this supercluster. The A3558 complex (A3558 
is the richest cluster of the region) is a structure elongated for $\sim 7$ 
\hmpc in the East-West direction, comprising also A3562, A3556, SC1329-313 
and SC1327-312. The A3528 complex extends for $\sim 7$ \hmpc along the 
North-South direction and is formed by two pairs (actually A3528 is double) 
of interacting clusters, including also A3530 and A3532. 

Dynamical studies of \cite{bardelli00} and \cite{reisen00} concluded that 
these structures (with mass of a few $10^{15}\ M_\odot$ each) are in the 
collapse phase, while the entire central supercluster region already reached 
its turn-around radius and has started to collapse.
In fact, the complexes represent a major merger at an advanced stage
(the A3558 complex) and at an early stage (the A3528 complex).
These two strucures are connected by a ``bridge" of galaxies, resembling the 
Great Wall: within this wall the overdensity in number of galaxies is 
$\sim 4$, consistent with the overall overdensity of $3.3$ obtained by 
\cite{drinkwater04} after having eliminated the cluster regions.

Formally, there is also another complex, dominated by A3571 (the cluster 
visible on the South-East in the lower panel of Figure 2) which is connected 
with A3572 and A3575 (two poor concentrations) in the optical image.
However, in the X-ray this cluster appears well relaxed and this could indicate
that the merging is ``old", in the sense that the gas had time to go to the
equilibrium, while galaxies are still at the end of the relaxation process.

\begin{figure}
\centering
\caption{{\it Upper panel:} distribution of optical galaxies to
$b_J$=19.5 in the central region of the Shapley Concentration. 
{\it Lower panel:} ROSAT-PSPC pointed observations mosaic of the same area. 
Note the presence of the two cluster complexes both in the optical
and the X-ray bands.}
\end{figure}

\begin{figure}
\centering
\caption{{\it Upper panel:} the A3558 cluster complex from the ROSAT-PSPC 
mosaic of \cite{ettori97}. {\it Lower panel:} ASCA hardness ratio map of the 
A3558 complex from \cite{akimoto}. Note the ``hot spot" between A3562 and
SC1329-313 (reproduced with the permission of the American Astronomical
Society).}
\end{figure}

\section{The A3558 complex}

Clusters belonging to this structure are embedded in a continuous envelope
of both hot gas (\cite{kull}) and galaxies 
(\cite[Bardelli et al. 1998a]{bardelli98a}) on a scale of $\sim 7$ \hmpc, 
i.e. surrounding the entire structure (see Figure 3). The hot gas has not had
origin from the cosmological filament (or ``wall") seen in the redshift survey, 
but probably it is intracluster gas expelled from the clusters by the merging 
(see the spatial analysis of A3562 in \cite[Ettori et al. 2000]{ettori00}). 
Also the galaxy envelope has had the same origin, being formed by the less 
bounded cluster objects, shared by the whole structure after the merging.

ROSAT, ASCA and Beppo-SAX (\cite[Bardelli et al. 2002]{bardelli02}, 
\cite[Hanami et al. 1999]{hanami}, \cite[Akimoto et al. 2004]{akimoto},
\cite[Ettori et al. 2000]{ettori00}) X-ray studies of this region did not 
detect shocks, although the gas distribution shows clear signs of disturbance. 
Only between A3562 and SC1329-313, in the Eastern part of the structure, 
a hotter region is detected (see Figure 3).
Furthermore, SC1329-313 has a gas distribution particularly disturbed with a
comet-like shape (\cite[Bardelli et al. 2002]{bardelli02}): an ASCA analysis 
of its X-ray spectrum led \cite{hanami} to claim the existence of significant 
turbulent motions or of a multiphase gas. This means that the merging is still 
at work and the clusters are far from equilibrium.
As can be seen from Figure 4, most of the gas distribution features in the 
A3558 complex could be related with galaxy distribution substructures. 

This spectacular major merging represents a unique opportunity to study
the effect of merging at radio wavelengths (see also the contributions of Zucca
and Giacintucci, this conference). In particular, a peculiar radio feature 
has been detected: it is formed by a radio halo and by a diffuse radiosource 
(see Figure 5); the only other known case is in the Coma cluster.
The radio halo, detected at the center of A3562, is nearby an head tail
radiogalaxy: we verified (\cite[Venturi et al. 2003]{venturi03}) that  
this radiogalaxy furnished the electrons which, after the reacceleration 
by a merging 0.4 Gyrs ago, are responsible for the halo emission. 
Also the radio spectrum of the halo is consistent with the last electron 
acceleration happened 0.4 Gyrs ago.
Moreover, we found that there is a significant lack of radiosources in
this structure (\cite[Venturi et al. 2000]{venturi00}): this signal is coming 
mainly from the cluster A3558 and could indicate that merging could switch-off, 
at least for a period, the radiosource activity.
Moreover, a relic radiosource has been found in the Westernmost part of the 
A3558 complex: a geometrical and dynamical reconstruction of this part of the 
structure lead to speculate that this relic had origin on the shock front 
(up to now undetected in the X-ray), caused by a small group infalling onto 
the A3556 cluster (\cite[Venturi et al. 1998]{venturi98}).
A3556 itself presents peculiar characteristics, because of its very low
X-ray surface brightness with respect to the optical richness: moreover,
its optical luminosity function presents an unusual shape, with a pronouced 
excess of bright galaxies (\cite[Bardelli et al. 1998a]{bardelli98a}).

\begin{figure}
\centering
\caption{{\it Left panel:} ASCA surface brightness distribution of the cluster 
A3558, from \cite{akimoto} (reproduced with the permission of the American 
Astronomical Society). Note the clumpy distribution.
{\it Right panel:} optical isodensity contours; different symbols correspond to 
galaxies belonging to different substructures found in the $\alpha$, 
$\delta$ and velocity space (see \cite[Bardelli et al. 1998b]{bardelli98b}). 
}
\end{figure}

\begin{figure}
\centering
\caption{{\it Upper panel:} VLA 20cm isophotes overplotted to the Digital Sky 
Survey, in the A3562 region. On the left the system radio halo plus 
head-tail radiogalaxy (not resolved) is visible. Note the peculiar radiosource 
on the right. {\it Lower left panel:} the head-tail radiogalaxy seen at higher 
resolution. This object furnished the electrons to the halo. 
{\it Lower right panel:} radio spectrum of the halo. Solid lines are models 
with different reacceleration times. }
\end{figure}

\begin{figure}
\centering
\caption{XMM-Newton observations of A3528. {\it Left panel:} X-ray surface
brightness distribution; note the hot bridge connecting the two
subclumps. {\it Middle panel:} temperature map. {\it Right panel:} surface 
brightness residuals, obtained after having subtracted a smoothed 
distribution.}
\end{figure}

\section{The A3528 complex}

A3528, the dominant cluster of the complex, is a double cluster formed
by two twins subclumps separated by 0.9 \hmpc and the other two clusters
of the complex (A3530 and A3532) are a close pair, separated by $\sim 1$
\hmpc. 

\cite{gastaldello} studied A3528 with XMM-Newton observations, obtaining 
surface brightness, temperature and abundance maps. Although a bridge of hot 
gas connecting the two clumps has been found, no shock is detected
(see Figure 6): this fact is unexpected, given the estimated masses of the 
clumps ($\sim 8\times 10^{13}\  M_\odot$ each) and their relative distance. 
The most reasonable explanation is that the merging was not head-on but
off-axis.
After having subtracted a $\beta$ model from the surface brightness of the 
two subclumps, we found emission excesses which can be used to determine 
the infalling direction (see right panel of Figure 6).
The conclusion is that this system is in an off-axis post-merging phase, 
with the closest core encounter happened $\sim 1-2$ Gyrs ago.
The interesting point is that the optical blue luminosities of the
two subclumps, which are twins for what regards the X-ray properties,
differ by an order of magnitude. This could indicated that one of the two 
clumps suffered more than the other of the galaxy ``pealing" process,
probably induced by a larger path through the large scale environment.
XMM-Newton data on the couple A3530/A3532 are presently in the reduction phase.

Our general conclusion is that, although the two single pairs of clusters
(the two clumps of A3528 and A3530/A3532) are mergings at an advanced state, 
the A3528 complex as a whole is at an earlier moment of collapse 
with respect to the A3558 complex, and the masses involved here are probably
lower.

\section{Summary}

Central regions of superclusters give the possibility
to find cluster mergings at different phases and strengths and to study
the consequences of cluster collisions on the intracluster medium and the 
galaxy population.  
The case study of the Shapley Concentration which I presented here shows
that the multiwavelength approach is the best way to analyse merging clusters. 

%



\end{document}